\documentclass[traditabstract]{aa}
\usepackage{graphicx}
\usepackage{txfonts}
\usepackage{natbib}
\usepackage{mathrsfs}

\newcommand{\ks}{km\,s$^{-1}$}
\newcommand{\vsin}{$v$~sin~$i$ }

\newcommand{\tef}{$T_{\rm eff}$ }
\newcommand{\lgg}{{\rm log}~$g$ }

\newcommand{\cd}{d$^{-1}$}

\newcommand {\logg}{\log g}
\newcommand {\corot}{CoRoT }
\newcommand {\eb}{CoRoT~102918586~}

\newcommand{\kms}{\mathrm{km\,s}^{-1} }
\newcommand{\teff}{$T_{\rm eff}$}

\begin{document}

   \title{CoRoT 102918586: a $\gamma$ Dor pulsator in a short period eccentric eclipsing binary. \thanks{Based on photometry collected by the CoRoT space mission and spectroscopy obtained with the Sandiford spectrograph attached at the 2.1m telescope at McDonald Observatory, USA and the FEROS spectrograph attached to the ESO2.2m telescope at ESO, La Silla, Chile. \
  The CoRoT space mission was developed and is operated by the French space agency CNES, with participation of ESA's RSSD and Science Programs, Austria, Belgium, Brazil, Germany and Spain}}


   \author{C. Maceroni\inst{1},
   J. Montalb\'{a}n\inst{2},
   D. Gandolfi \inst{3,4},
   K. Pavlovski \inst{5},
\and
   M. Rainer \inst{6}
          }

   \institute{INAF--Osservatorio astronomico di Roma, via Frascati 33, I-00040 Monteporzio C., Italy\\
              \email{maceroni@oa-roma.inaf.it}
         \and
             Institut d'Astrophysique et G\'{e}ophysique Universit\'{e} de Li\`{e}ge,
		All\'{e}e du 6 A\^{out}, B-4000 Li\`{e}ge, Belgium. 
	\and 
	ESA Estec, Keplerlaan 1,  2201 AZ Noordwijk, Netherlands
	\and 
	Th\"{u}ringer Landessternwarte Tautenburg, Sternwarte 5, D-07778 Tautenburg, Germany.
	\and
	Department of Physics, University of Zagreb, Bijeni\v{c}ka cesta 32, 10000 Zagreb, Croatia.
	\and
	INAF, Osservatorio Astronomico di Brera, Via E. Bianchi 46, I-23807, Merate, Italy.
             }

   \date{Received November 16, 2012; accepted January 30, 2013}

\authorrunning{C. Maceroni et al.}

 
  \abstract
   {Pulsating stars in eclipsing binary systems are  powerful tools to test stellar models.  Binarity enables to constrain  the pulsating component physical parameters, 
  whose knowledge drastically improves the input physics for asteroseismic studies. The study of stellar oscillations allows us, in its turn, to improve our understanding of stellar interiors and evolution. The space mission CoRoT  discovered several promising objects suitable for these studies, which have been photometrically observed with unprecedented accuracy, but  needed spectroscopic follow-up. A promising target was the relatively bright eclipsing system  CoRoT~102918586, which turned out to be a double-lined spectroscopic binary and  showed, as well, clear evidence of $\gamma$ Dor type pulsations. 
  
With the aim of combining the information from binarity and pulsation and fully exploit the potential of CoRoT photometry we
  obtained  phase resolved high-resolution spectroscopy  with the  Sandiford spectrograph at the McDonald 2.1m telescope and the FEROS spectrograph  at the ESO 2.2m telescope. Spectroscopy yielded both the  radial velocity curves and, after spectra disentangling, the component effective temperatures,  metallicity and line-of-sight projected rotational velocities.   The CoRoT light curve was  analyzed with  an iterative procedure, devised to disentangle  eclipses  from pulsations.  The eclipsing binary light curve analysis, combined to the spectroscopic results, provided an accurate determination of the system parameters, and the comparison with evolutionary models strict constraints
 on the system age.
   Finally, the residuals obtained after subtraction of the best fitting eclipsing binary model  were analyzed to determine the pulsator properties.
  
   We  achieved a quite complete and consistent description of the system. The primary star pulsates with typical $\gamma$ Dor frequencies and  shows a splitting in period which is consistent with high order g-mode pulsations in a star of the corresponding  physical parameters. The value of the splitting, in particular,  is consistent with pulsations in  $\ell=1$ modes.
   
   }


   \maketitle
%

\section{Introduction}
In the last six years the \corot space mission has   acquired more than 150,000 high-accuracy light curves with dense and almost continuous sampling  over time intervals from one to five months.  While the main goals of \corot are exoplanet hunting by the transit technique and asteroseismology of bright stars, a precious by-product is a large number of variable stars, most of them new discoveries.  

Eclipsing binaries (EBs)  outnumber all other variables found by \corot and  some among them reveal additional regular variability superimposed on the binary light curve.  The origin of the additional variability  is very often rotation modulated  surface activity,  in some cases -- however --  there is  clear evidence of pulsations, including the  non-radial pulsations typical of $\delta$ Sct, $\gamma$ Dor and Slowly Pulsating B stars (SPB).

  Precious insights in stellar structures can be obtained from asteroseismology, and pulsating EBs have a fundamental asset: studying non-radial oscillations in  EB components has the advantage that the masses and radii  can be independently derived,  with a pure geometrical method, by combining the information from the light and and the radial velocity  curves, and with uncertainties, in the best cases, below 1\% \citep[e.g.,][]{South05}. Moreover, additional useful constraints derive from the requirement of same chemical composition and age.
  As the precise measurement of the mass and radius  poses  powerful constraints on  the pulsational properties,  EBs with pulsating component potentially provide direct tests of the modeling of complex dynamical processes occurring in stellar interiors (such as mixing length, convective overshooting, diffusion). 
Furthermore, a close eclipsing binary containing a pulsating component is the ideal laboratory to study the effect of tides on stellar pulsations.

The trade off with these advantages  is in a  more complex structure and analysis of the data, as it is necessary to disentangle  the  two phenomena at the origin of the observed time series, but the results for the first \corot targets of these type studied so far \cite[e.g.,][]{cm09, dam10, sokoletal10} are quite encouraging.
 
Most EBs  observed by CoRoT belong to the exoplanet field, whose targets are in the V--magnitude range   11-16.5. Each CoRoT run provides the light curves of several thousands exo-targets (up to twelve thousands for the first runs and about half that number  after the loss of one of the two data processing units in March '09). The variable stars are automatically classified by the CoRoT Variability Classifier \citep{cvc09} which yields the membership probability for 29 different variability classes.  

The detailed inspection of the light curves of  EBs and  of pulsating stars of the first CoRoT runs allowed us to select a small number of variables showing both eclipses and oscillations. All of them were new discoveries and the only available spectroscopic measurements (if any) were low-dispersion spectra from a ground based survey of \corot fields (see next section). 
We  organized, therefore, a  program aimed to obtain phase resolved echelle  spectroscopy of the suitable targets in the sample.

In this paper we present the results relative to the brightest object of our sample: CoRoT~102918586, which was classified as EB and $\gamma$ Dor candidate. 

 The  $\gamma$ Dor variables are A-F stars of luminosity class IV-V  pulsating in high-order gravity-modes, with typical frequencies  in the range 0.3-3 days.  The driving mechanism of pulsations is  the modulation of the radiative flux by convection at the base of the convective envelope \citep{guz00, mad05}. The g-modes  probe  the deep stellar interior, and - for this reason - are of great interest for asteroseismic studies. 

Before the CoRoT and {\it Kepler} space missions  sixty-six $\gamma$~Dor variables were  known, according to  \cite{hf07}, and about 50 \% of them were found in binary stars
(12 visual binaries, 10 double-lined  and 6 single-lined spectroscopic binaries, no EB). The first $\gamma$ Dor in an eclipsing binary was announced by \cite{iba07} but the quality and the coverage  of their ground-based photometry was insufficient for a detailed asteroseismic study. Two new candidate $\gamma$ Dor's  in EBs were identified by \cite{dam10} and \cite{sokoletal10} (CoRoT~102931335 and CoRoT~102980178) but their faintness prevented the collection of spectroscopic data and, therefore, the full exploitation of the  EB assets. 
Many others similar objects have been later identified in the CoRoT and {\it Kepler} archives, but again, without spectroscopic information.

 \eb is, therefore, the first $\gamma$ Dor in an EB which can be studied in great detail, thanks to the quality of the CoRoT photometry,  to its brightness -- making possible to  acquire high dispersion spectroscopy -- and to  the  fortunate occurrence of being a double-lined spectroscopic binary.

This paper is organized as follows: Section \ref{phot} and \ref{spectro} describe the available data and their reduction, Section 
\ref{lc-rv-an}  deals with the light and radial velocity curve analysis, Section \ref{abundances} is devoted to the analysis of the disentangled component spectra, providing the atmospheric parameters,  Section \ref{abspar}  presents a  comparison between the physical elements derived from the analysis and stellar evolutionary models and, finally,  Section \ref{puls}  analyzes  the pulsational properties of the primary component.


\section{CoRoT photometry}
\label{phot}

\eb is one of the $\sim$ 10000 exoplanet targets observed by \corot during the ``Initial Run" (IRa01), the first scientific run   which lasted for about $60^{\mathrm{d}}$.
The little pre-launch information  on this target was collected in Exo-dat database  \citep[operated at CeSAM/LAM, Marseille, France  on behalf of the CoRoT/Exoplanet program, ][]
{deleuil09}. The target is relatively bright ($V=12.45 \pm 0.03$). Prior to CoRoT observations its binary nature was unknown.  We derived a spectral classification F0~V by
carefully  re-analyzing \citep{cmlanz}   low-resolution ($R \approx 1300$)  spectra, which were  obtained with the AAOmega multi-object facilities of the Anglo-Australian Observatory, in the frame of a program aimed to acquire a first spectroscopic snap-shot of the CoRoT fields \citep{aaom1, aaom2}. This spectral classification is excellent agreement  with  the results obtained, later on,  from the  Sandiford and FEROS  observations. 

  The point-spread function (PSF) on the CoRoT exoplanet field detectors is actually a mini-spectrum (R$\sim$3), thanks to an objective prism in the optical path of the CoRoT exoplanet channel. Three-colour photometry (the so called red, green, and blue colours) is obtained by splitting the PSF into three sub-integrating areas for target stars brighter than $V\le15.2$. Those colours do not correspond to any standard photometric systems, though \citep{auvergne09}. Yet, they are useful to photometrically rule out false positives mimicking a planetary transit signal \citep{carone12}. Therefore, even if chromatic light curves were available for our target, our analysis is based mainly on the white light curve, which was obtained summing up all the channels.

\begin{figure}[ht!]
   \centering
   \includegraphics[width=9.cm]{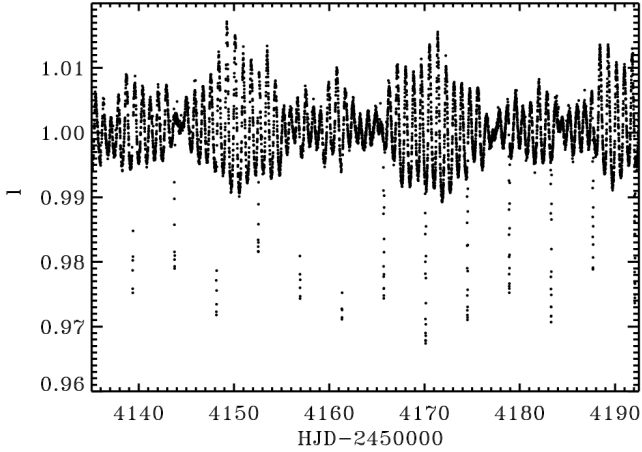}
      \caption{\eb de-trended white-light light-curve after  rebinning  to 512$^{\mathrm{s}}$ and normalization to the mean value (eclipses excluded).}
         \label{lc}
   \end{figure}

This original light curve contains $\sim 78000$ points, with a first section (HJD $< 2454162^{\mathrm{d}}$)  sampled in the long integration mode (512$^{\mathrm{s}}$)
and the rest in the short one (32$^{\mathrm{s}}$). Having checked the absence of high frequency components, we rebinned the time series to the longer  step and corrected a long term trend,  dividing the light curve by  a third order Legendre polynomial. Besides, we discarded with a sigma-clip algorithm the obvious outliers. 
The result of these preliminary operations is shown in Fig.~\ref{lc},  the total number of points is  reduced to  8062.  Hereafter we will refer to this curve as LC0.

The first step of our iterative procedure was the pre-whitening of LC0,   with the result of the harmonic fit obtained with Period04  \citep{p04}, i.e. a   non-linear least square fit of the frequency, amplitude and phase of the simusoidal components, as derived from the Discrete Fourier Transform (DFT) of the time series. Period04 was applied to LC0 after  eclipse masking.
 The harmonic fit  was obtained with the  twenty-four frequencies  having an amplitude   S/N$>$4, according to the criterion proposed by  \cite{breger93}. The first thirteen frequencies, in order of amplitude, are listed in Table~\ref{hf0} together with corresponding amplitudes, phases  and S/N ratios.
 
 In this type of analysis of a single light curve  a problem often arises concerning the nature of the orbital period harmonics. On one side, in fact, there could be pulsations excited at orbital overtones by tidal action. On the other  features that belong to the orbital variation (e.g. out of eclipse variations due to proximity effect or eccentricity)  produce orbital overtones in the frequency spectrum.  Besides any deviation of the binary model from the real configuration generates orbital overtones in the residuals obtained by subtraction 
 of the binary model. The analysis of \eb
 is a good example in this respect: the first inspection of the light curve, and the knowledge about the luminosity class of the primary star, suggested a binary configuration with two components of similar surface brightness (and hence almost equally deep minima) in a circular orbit, and with  an orbital period of $\approx 8\fd78$. This implies, as well, small fractional radii of the components and very little  tidal deformation. 
 
While it was possible to obtain an excellent fit of binary + pulsations with this model \citep{cmlanz}, it was problematic to explain the presence of several orbital overtones in the 
harmonic fit (see Table~\ref{hf0}), as tidal forces should be weak in  {\em spherical} components  describing  {\em circular} orbits.  

The alternative hypothesis, suggested in the above mentioned paper, was a completely different  system configuration: the minima seen in the curve are the same, single  eclipse. In that case the system should be formed either by  very different stars,  or have an elliptic orbit with   such an orientation in space to produce an eclipse only when the components are close to periastron passage  \citep[e.g., a case similar to that of   HD~174884, ][]{cm09}.
This second hypothesis resulted to be the true one after collection of the first series of spectra: the orbit turned out to be  elliptic,  with half the orbital period, and its inclination angle and longitude of periastron yield a single, grazing, eclipse (see Section \ref{spectro}).

\begin{table}[thb!]
\caption{Results of the preliminary harmonic analysis.}
\label{hf0}
\begin{tabular}[ht!]{llllc}
\hline\hline
F (\cd)& Ampl $\cdot 10 ^3$ &Phase (2$\pi$ rad) & S/N$^a$ &remark \\ 
\hline
	1.22452 	(2)&	4.30116 (1)&	 0.1576 (3)	&24.7&\\
	1.12666 	(3)&	3.05882 (1)&	 0.1223 (4)	&21.7&\\
	1.17256 	(3)&	2.64945 (1)&	 0.6155 (5)	&24.2&\\
	0.94653 	(4)&	1.43687 (1)&	 0.2809 (9)	&16.5&\\
	2.3512 	(1)&	0.72474 (1)&	 0.101   (2)	&11.8&\\
	0.4572 	(1)&	0.66486 (1)&	 0.432   (2)	&8.51& 2 F$_{\mathrm{orb}}$\\
	0.2277 	(1)&	0.53671 (1)&	 0.286   (2)	&7.00&  ~~~F$_{\mathrm{orb}}$ \\
	2.39622 	(1)&	0.49724 (1)&	 0.903   (3)	&9.77&\\
	0.0519 	(2)&	0.50943 (1)&	 0.251   (3)	&6.72&\\
	2.4490 (2)& 0.41644 (1)&	 0.097   (3) 	&9.65&\\
	0.6833 (2)& 0.44263 (1)&	 0.055   (3) 	&7.96& 3 F$_{\mathrm{orb}}$\\
	2.3009 (2)& 0.42179 (1)&	 0.944   (3)	&9.33&\\
	0.9118 (2)& 0.32531 (1)&	 0.901   (4)	&6.49& 4 F$_{\mathrm{orb}}$\\
\hline 
\end{tabular}
\begin{list}{}{}
\item[] {Number in parenthesis: formal errors on the last digit of the LS fit.}
\item[$^{\mathrm{a}}$] {S/N is computed over an interval of 5 \cd}
\end{list}
\end{table}
   \begin{figure}
   \centering
   \includegraphics[width=9.cm]{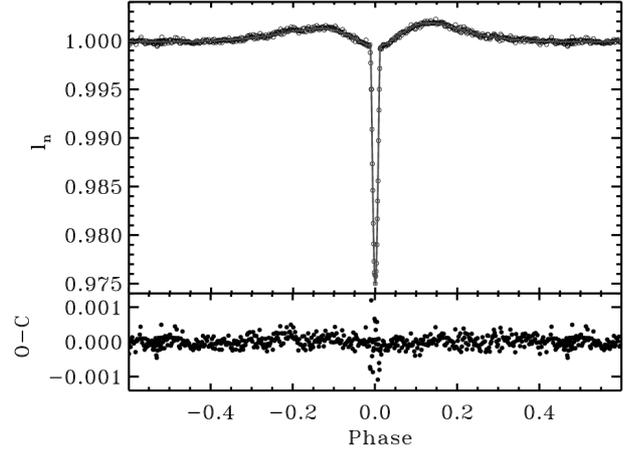}
         \caption{Upper panel: the phased light curve of the eclipsing binary as  residuals of the harmonic fit, orbital overtones excluded.   Full line:   final fit with PHOEBE.  Lower panel: fit residuals. The displayed curve is the final result of our procedure (after two steps of iterative pre-whitenening, see text). }
         \label{lceb}
   \end{figure}
The contribution to LC0 only due to eclipses (i. e. the residuals of subtraction of the harmonic fit), was then recomputed subtracting  only the 20 frequencies not corresponding to orbital overtones. This orbit-only light curve (Fig.~\ref{lceb}, LCEB)  is no longer flat out of eclipse  and shows two almost equal bumps on the sides of the single minimum, which is exactly the expected shape  for a longitude of periastron close to 90 degrees,  i.e. a line of view coplanar with  the system semi-axis.

The light curve of Fig.~\ref{lceb} is phased with the ephemeris:  
\begin{equation}
  T_{\mathrm{minI}} = 2454139.3798 (4) +   4 \fd 39138 (6) \times E.
\label{ephe}
\end{equation}
which was derived by a linear fit to the minimum times,  computed in their turn with a parabolic fit to the lower section of each minimum. The figure also shows the synthetic light curve of the best fitting binary model computed  with PHOEBE \citep{prsazw05} and described in Section \ref{lc-rv-an}.

 For the sake of brevity we directly show   the final result of the iterative process used to extract LCEB  from LC0. The procedure yielding the final LCEB is:
 \begin{enumerate}
 \item prewhitening of LC0 with 20 out of the 24 frequencies obtained in the first harmonic fit, as already described
 \item phasing with the ephemeris of  Eq. \ref{ephe} and further binning with a variable step, to limit computing time for PHOEBE fitting (namely computation of normal points  in phase bins of 0.001 and 0.002 in and out of eclipse).
\item subtraction of the PHOEBE fit model  from LC0 and new harmonic fit of this full time series.
\item prewhitening of LC0 with the results of the  previous step, yielding a new 'orbit-only' light curve.
\end{enumerate}
The procedure was stopped after this last stage because there was no significant change in  the parameters of the binary model and in the results of the harmonic fit.

 The resulting LCEB,which is phased and binned as in step 2,  is shown   in Fig.~\ref{lceb}, together with  the residuals  of the PHOEBE fit. Note that
 the larger scatter of residuals at primary minimum, and at the predicted position of the secondary one, is partly due to the smaller number of  points in the variable width bins. 
\section{High resolution spectroscopy of \eb} 
\label{spectro}
\eb is the brightest target of the sample of \corot binaries with pulsating components we selected for spectroscopic follow-up. We collected echelle spectra with two different instruments, namely:   the Sandiford spectrograph, attached on the Cassegrain focus of the 2.1\,m Otto Struve telescope at McDonald Observatory (Texas, USA) and
the fiber-fed FEROS spectrograph  mounted on the ESO 2.2m telescope in La Silla, Chile.

Seven Sandiford spectra were acquired in December 2009 and three in January 2011. We used a $1\,\arcsec$ slit, which gives a resolving power of $R=47\,000$. The small size of the CCD detector enables to cover only $\sim1000\,\AA$. We thus set the spectrograph's grating angle to encompass the $5050-6030\,\AA$ wavelength spectral region, which is a good compromise between obtained signal-to-noise (S/N) ratio and number of spectral lines suitable for radial velocity (RV) measurements. Two consecutive spectra of 20-30 minutes were secured in each observing night and subsequently co-added to remove cosmic rays hits.  The inspection of the first seven  spectra, acquired at McDonald in December 2009, revealed that \eb is a double-lined spectroscopic binary with an orbital period of about 4.39 days.

Seven more spectra were taken with FEROS in December 2010. FEROS is fed with a $2\,\arcsec$ fibre, which yields a resolution of $R=48\,000$ and a spectral coverage from 3800 to 8500\,$\AA$. We adopted the same observing strategy as that used in the Sandiford observations. 

 The reduction process involved the usual steps of de-biasing, flat-fielding, background subtraction and wavelength calibration by means of measurements of a ThAr calibration lamp. The spectral section of interest were then normalized to the continuum and barycentric corrections were computed for the time of mid-exposure. The FEROS spectra had a typical S/N of 50-60 in the  region 5500-5800 \AA\, and of 30-40 at 4800 \AA, while Sandiford  spectra have a typical S/N 40-50 at 5500 \AA.

The component radial velocities were derived with a  standard cross correlation algorithm, IRAF's tool FXCOR. In the case of FEROS spectra, we used the wavelength interval
4500 -- 4800  \AA, rich of metallic lines. As cross correlation  template, in the FEROS case,  we chose a nearby  bright CoRoT primary target (HD 49933) of similar spectra type  (F2) and we used, for reference, the RV standard HD~22484. The cross correlation template used for the Sandiford spectra was instead the spectroscopic standard HD\,50692 \citep{nidever02}.  The  radial velocities of the components were derived by  fitting two Gaussian curves (with the FXCOR ``deblend" function),  obviously the fit errors depend on the
separation of the CCF peaks.

\begin{figure}[htb]
   \centering
   \includegraphics[width=8.5cm]{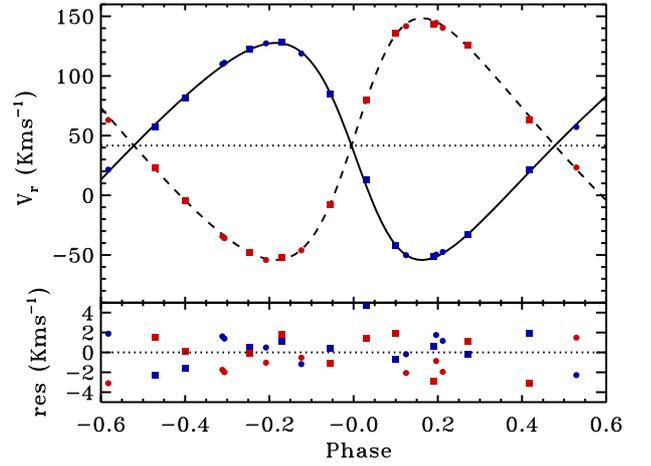}
         \caption{Upper panel: the phased radial velocity curves of  \eb components. Filled circles: FEROS measurements, filled squares Sandiford data; the uncertainties are smaller than the symbol size.
      Lines: PHOEBE fits (full: primary, broken: secondary component).  Lower panel: fit residuals, with the same symbols. }
         \label{rvc}
   \end{figure}

The radial velocity curves, showing excellent agreement between the two data sets, are shown in Fig.~\ref{rvc}, the corresponding values in Table~\ref{rv}. 
 Both Sandiford and FEROS uncertainties include those of the RV standard stars \citep[0.1\ks;][]{nidever02, rvst},
the FEROS data (the last seven rows in Table~\ref{rv})  include as well  the uncertainty on the template velocity (0.1 \ks).
 \begin{table}
 \caption{Radial velocities of CoRoT~102918586}    
\label{rv}      
\centering                          
\begin{tabular}{c r r c}        
\hline\hline                 
BJD -2450000 &  $v_{\mathrm{rad,1}}$  &  $v_{\mathrm{rad,2}}$ & Instrument$^a$\\    
                      &    (\ks)     & (\ks)   & \\
\hline                        
5173.91504  &  81.6 $\pm  0.4 $ &  -4.5  $\pm 0.5$ & S \\
5174.92337  & 128.4 $\pm 0.3  $& -51.9  $\pm 0.3$ & S \\
5175.79674  &  13.3 $\pm  0.4$ &  79.9  $\pm 0.4$  & S\\
5176.85915  & -32.9 $\pm  0.2$& 125.5  $\pm 0.3$  & S\\
5177.99141  &  57.2 $\pm  1.0$ &  23.3  $\pm 1.0$  & S\\
5178.97184  & 122.5 $\pm 0.3$& -47.9  $\pm 0.3$  & S\\
5179.82042  &  85.3 $\pm 0.3$ &  -7.5  $\pm 0.4$  & S\\
5584.90412  & -51.6 $\pm 0.3$ & 143.3  $\pm 0.3$  & S\\
5585.90005  &  21.5 $\pm 1.0$ &  63.1  $\pm 1.0$  & S\\
5588.89373  & -42.1 $\pm 0.3$ & 136.1  $\pm 0.3$  & S\\
5556.80513  & 127.3 $\pm 0.5$ & -54.2  $\pm 1.0$ & F\\
5558.57947  & -49.7 $\pm $ 0.6& 144.5  $\pm 1.2$ & F\\
5558.64812  & -47.5 $\pm $ 0.5& 140.3  $\pm 1.1$ & F\\
5560.74155  & 110.0 $\pm $ 0.6& -34.4  $\pm 1.1$ & F\\
5560.76300  & 111.0 $\pm $ 0.5& -36.0  $\pm 1.2$ & F\\
5561.56488  & 118.8 $\pm $ 0.5& -46.1  $\pm 1.0$ & F\\
5562.65659  & -50.2 $\pm $ 0.6& 141.7  $\pm 1.1$ & F\\

\hline                                   
\end{tabular}
\begin{list}{}{}
\item[$^{\mathrm{a}}$] {S for Sandiford spectrograph and F for FEROS.}
\end{list}

\end{table}
\section{Light  and radial velocity curve analysis}   
\label{lc-rv-an}
\subsection{Determination of the  best fitting binary model.}
The light and radial velocity curve solutions were performed with the current (``devel") version of PHOEBE,   which  includes  flux computation with the CoRoT transmission functions, for both the seismology and the exoplanet field  \citep[for details see ][]{cm09}. 

We preferred to avoid the simultaneous solution of light and radial velocity curves, being the RV data acquired at different epochs with respect to photometry and being the data sets very different in terms of observed point number and accuracy. On the other hand the two solutions were connected by keeping fixed in each of them the parameters better determined by the other type of data.

In the light curve solution  we  adjusted  the inclination $i$,  the secondary effective temperature, $T_{\mathrm{eff,2}}$, and the surface potentials $\Omega_{1,2}$; the primary passband luminosity $L_1$ was separately computed  rather than adjusted,  as this allows for a smoother convergence to the minimum. The eccentricity $e$, mass ratio $q$, and the longitude of periastron $\omega$ were instead  fixed to the values derived  from the fit of the radial velocity curves (where we adjusted as well the system semi-axis, $a$, and the  barycentric velocity $\gamma$).

 For limb darkening we adopted a square root law that employs two coefficients, $x_{LD}$ and $y_{LD}$ per star and per passband. The coefficients are determined by  PHOEBE interpolating, for the  given atmospheric parameters, limb darkening coefficient tables. These tables  were computed by synthesizing SEDs for many  passbands and at different  emergent angle, and fitting  the linear cosine law and non-linear log and square root laws by least squares \citep[see ][]{cm09}.     
 The gravity darkening and albedo coefficients were kept fixed at their theoretical values, $g = 1.0$ and $A = 1.0$ for radiative envelopes \citep{vonzeipel1924}. 

The exoplanet fields were selected with the purpose of maximizing the number of late type dwarfs, which, thanks to higher contrast, are better suited for transiting planet detection. As a consequence the exo-fields are usually relatively crowded. Moreover the target PSFs, as we have already mentioned, are enlarged by the miniprism. To limit contamination each target flux is recorded only in a specific (software) mask, whose size and form depends on the star characteristics. Nevertheless contamination by faint unresolved targets is rather common. Information on the degree of contamination has been independently acquired by ground-based photometry programs with a much higher space resolution \citep{deeg09}.
For our system the contamination is non-negligible, being estimated to 15\% of the total flux (Deeg 2012, private communication),  a value similar to that appearing in the Exodat database (14.4 \%). We added, therefore, the former value as ``third light".

The primary effective temperature was kept fixed, as it is well known that the solution of a single light curve is sensitive only to relative values of the effective temperatures.  We started with a value of 7200 K (an estimate from the spectral type) and finally used 7400 K (suggested later on by the analysis of the disentangled spectra). As expected the change of T$_{\mathrm{eff,1}}$ reflected into a similar shift for T$_{\mathrm{eff,2}}$.  
The analysis of the line profiles (see section \ref{abundances}) suggested rotational velocities very close to spin orbit synchronization for both components, we adopted, therefore, a ratio of rotational to orbital period $f_{1,2}$=1. 

The system model corresponding to the best fit is formed by two almost twin  stars; the low inclination,  the eccentricity and the orientation of the orbit are at the origin of the single minimum, which is a grazing eclipse of the larger star.  The system parameters from the light and radial velocity curve solution are collected in Table~\ref{sol}. The unicity of the solution and the derivation of parameter uncertainties are discussed in the next section.

\begin{table}[tb]
\caption{Parameters from PHOEBE fits of light and RV curves and  derived physical parameters of CoRoT~102918586}
\label{sol}
\centering
\begin{tabular}{lccc}
\hline\
			&					& System		&                      \\
                    			& Primary	&						& Secondary     \\
\hline
$i$ (${}^\circ$)			&	 			& $ 77.66 \pm0.07$  			&			\\
$e$			   		&				&$0.249 \pm0.005$			&			\\
$\omega$				&				& $102.6^\circ  \pm 1.4	$	&			\\
$q$					&				&$0.898	\pm 0.007	$		&			\\
$a~ (R_{\odot})$		&				&$	16.53 \pm 0.07$		&			\\
$\gamma~ (\kms) $             &				&$42.0	 \pm 0.4$			&			\\
$(L_2/L_1)_{\mathrm{CoRoT}}$  			&				&$	0.673 \pm 0.015$		&			\\
$T_\mathrm{eff}$ (K)		& $7400^{a }\pm 90$	&  		& 7144$ \pm 110$		\\
$M  (M_{\odot})$                	&  $1.66 \pm0.02$		&                & $1.49\pm0.03$	\\
$R  (R_{\odot})$                	&  $1.64 \pm0.01$	         &                & $1.48\pm0.01$	\\
$\logg$                		&  $4.23 \pm0.01$	 	&                & $4.27\pm0.01$	\\
\hline
\end{tabular}
\begin{list}{}{}
\item[$^{\mathrm{a}}$] { Fixed value, the uncertainty is  from the analysis of Section~\ref{temps}.}
\end{list}

\end{table}

   \begin{figure*}[htb]
   \centering
   \includegraphics[width=8.5cm]{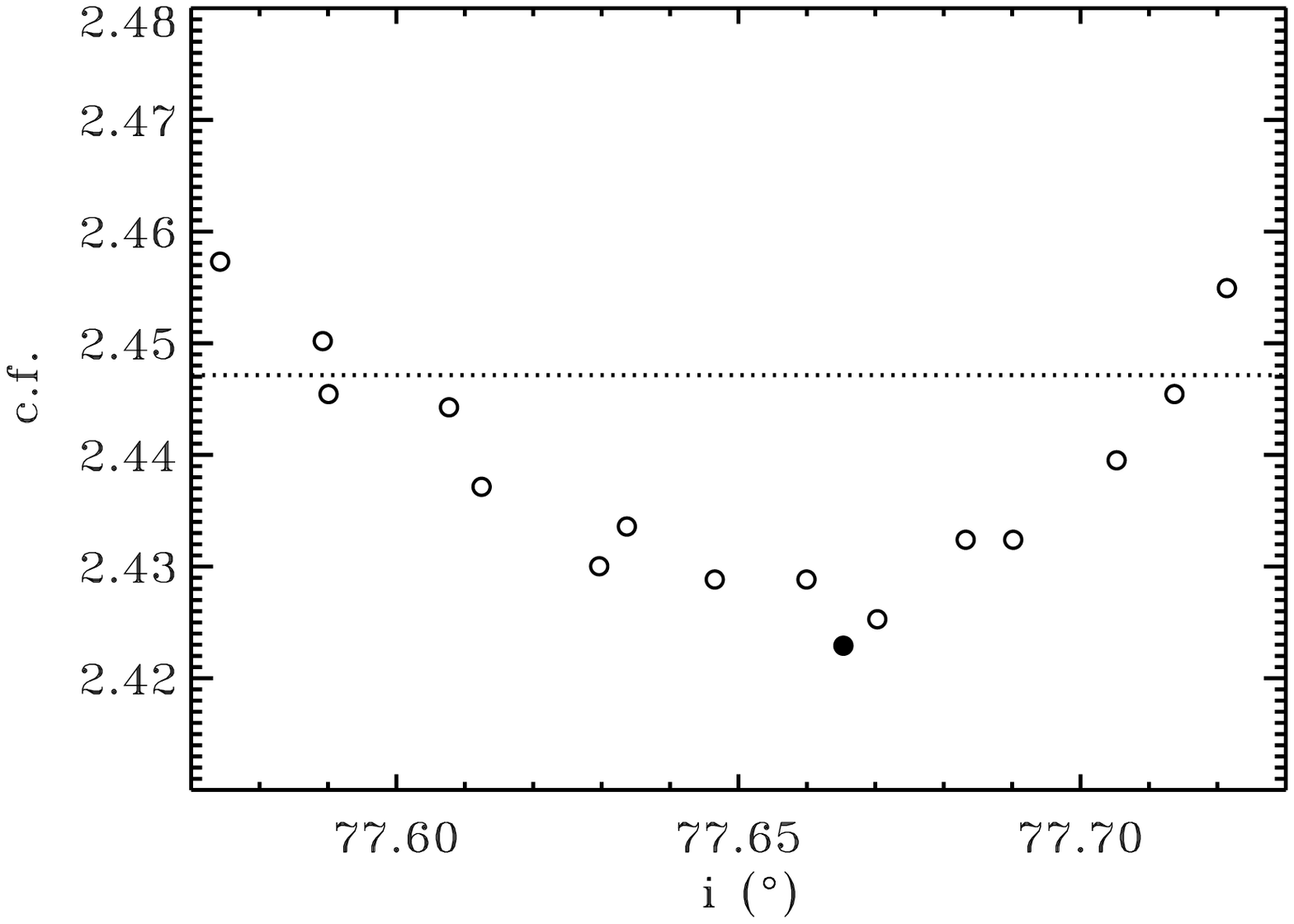}
   \includegraphics[width=8.5cm]{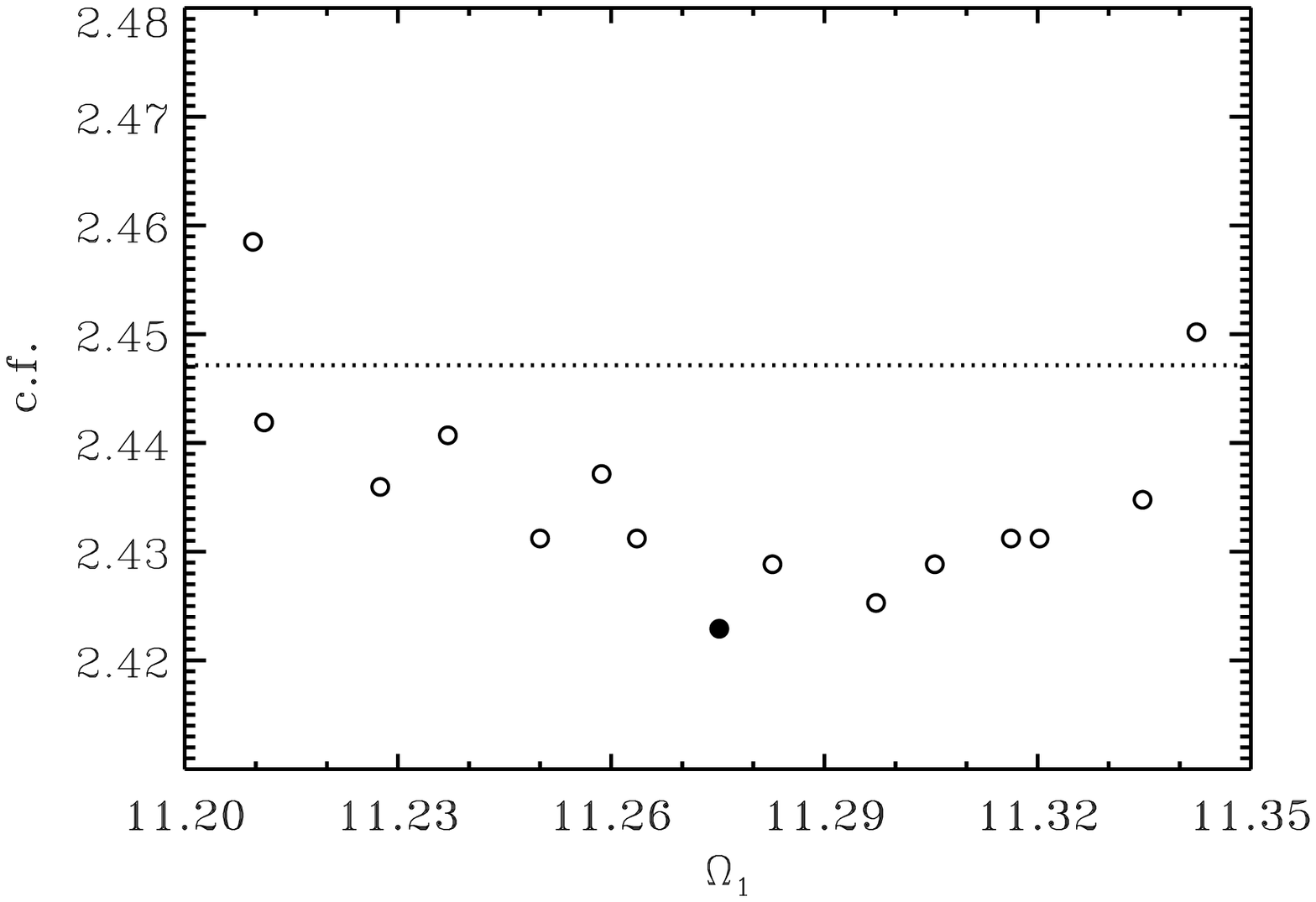}
   \includegraphics[width=8.5cm]{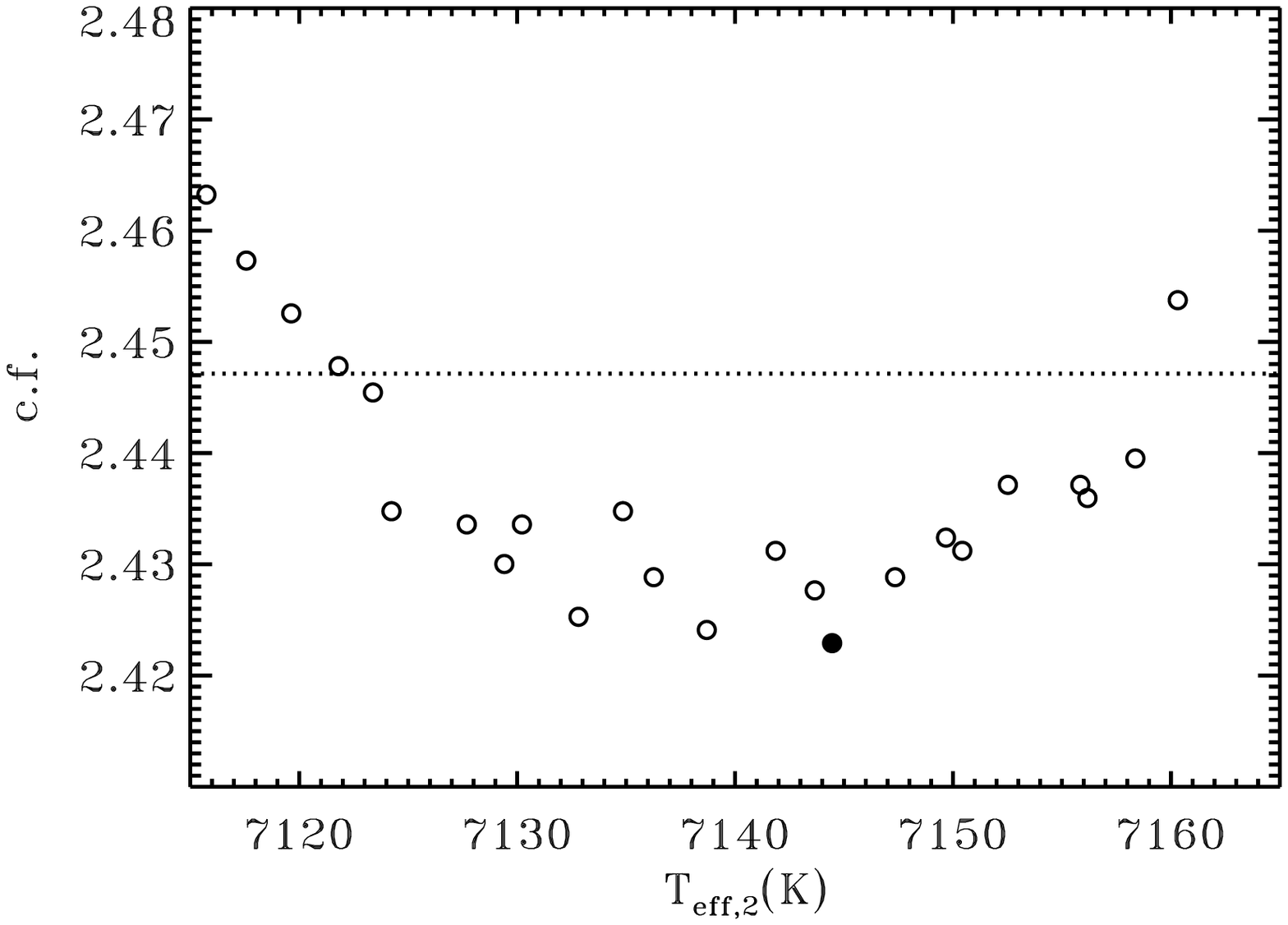}
   \includegraphics[width=8.5cm]{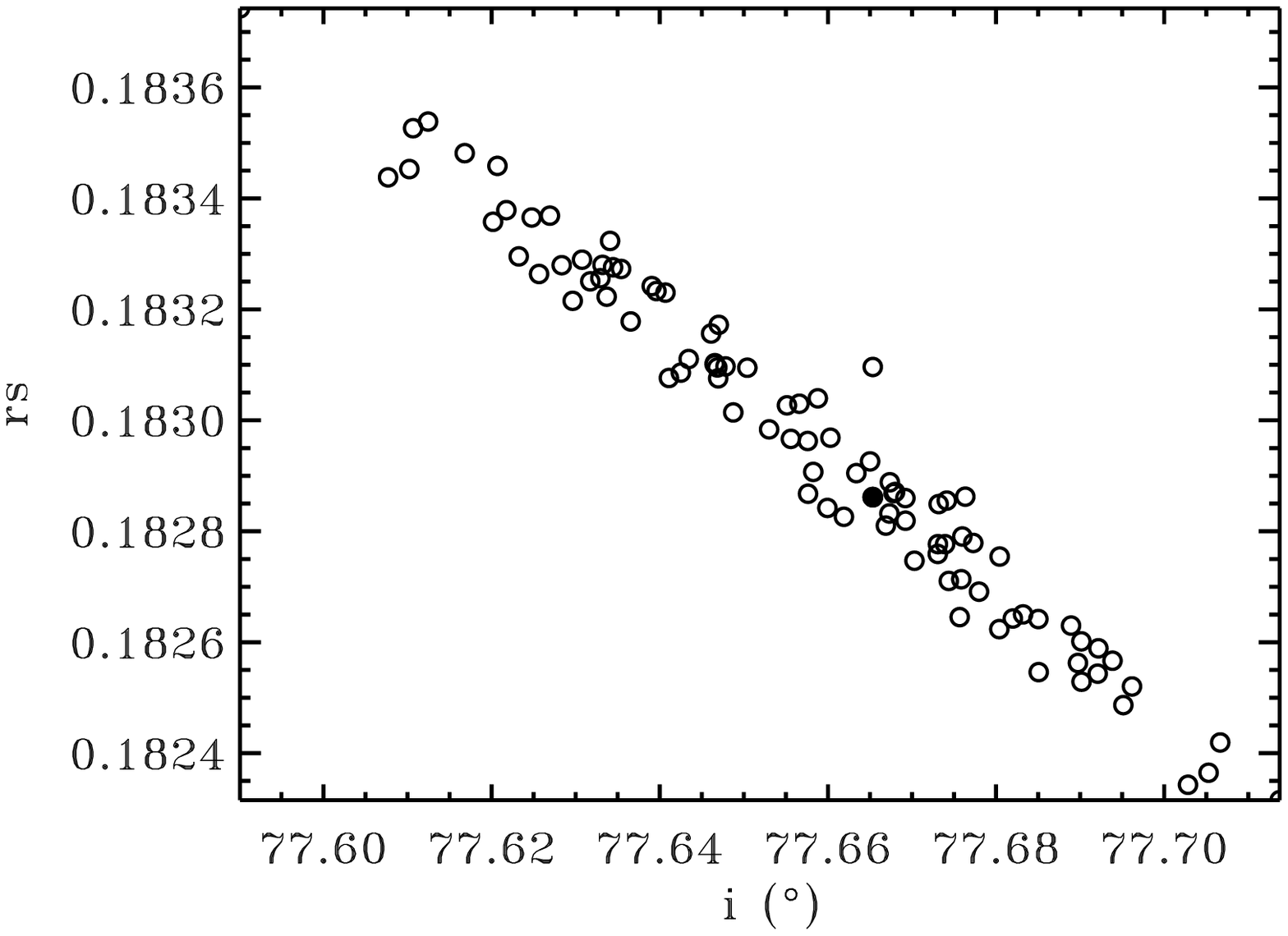}
         \caption{The projection of the cost function along   inclination (top left), primary surface potential (top right) and secondary temperature (bottom, left). The minimum value is
         indicated by a filled dot.
         Bottom right panel: the correlation of  $rs$ with inclination for the best fitting models (cost function value $\le$ 1 \% above the minimum). }
         \label{scans}
   \end{figure*}

\subsection{Unicity of the solution and parameter uncertainties} 
\label{uniq}
The light curve fit is based on the minimization procedure of a cost function measuring the deviation between model and observations in the space of adjusted parameters  ( in  PHOEBE differential corrections in the Levenberg-Marquart's variant or Nelder and Mead's downhill simplex). As a consequence it suffers of the well know problems of these methods: the minimization algorithm can be trapped in a local minimum or  degeneracy among parameters and data noise 
can transform the minimum into a large and flat bottomed region,  or an elongated flat valley, rather than a single point. 
Besides, the correlation among the parameters implies uncertainties on the derived values which are 
significantly larger than the formal errors derived, for instance, by the least square minimization. 
In the specific case of a detached binary with almost grazing eclipses one expects, for instance,   a strong degeneracy among fractional radii (i.e. surface potentials) and inclination. 

  To handle these problems we  performed an heuristic scan of the hyper-surface describing the cost function value in the space of the adjusted parameters.
  Minimizations were first performed starting from random assigned input points, allowing to locate the region(s) of minima.  The lowest minimum region was then further explored by mapping the cost function at randomly chosen points. Fig.~\ref{scans} displays the  lower envelope of the projections  on   different
 parameter plans  of the cost-function hyper-surface. 
 
 We chose, as uncertainty on the parameters, the interval corresponding to a variation of 1\% of the absolute minimum value. 
  The choice of  this threshold value is based  on the comparison with the results of  an independent estimate of uncertainties from  bootstrap resampling (BR), a very useful technique to estimate parameter confidence levels of the least squares solutions \citep[see, for instance,][]{NR}. 
 In short, BR consists in generating many different data samples by random resampling with repetitions (bootstrapping) the available data, performing the minimization procedure for each sample, and deriving confidence intervals from the resulting distribution of parameters. 
The interval of a given parameter containing 68.3\% of the solutions has the same meaning of   the 1-$\sigma$ interval of the Gaussian distribution.
The estimate by BR was performed both for the radial and the light curve solution, according to the scheme described in detail in \citet{mr97} and \citet{cm09}. The main point is that  the procedure is performed within the minimum already established by a single iterated solution (that is, using only one set of residuals and parameter derivatives).  The  68.3\% interval for the inclination from BR  is $\pm 0.07$. A comparison   the first panel of Fig. \ref{scans}  allows to derive the threshold value of 1\% used in the following.

  To illustrate the correlation between the sum of radii and the inclination we use the  quantity:
  \begin{equation}
  \label{rsum}
  rs=\frac{1}{\Omega_1-q}+\frac{q}{\Omega_2-0.5+0.5q}
  \end{equation}
where  $\Omega_1$, $\Omega_2$  are the surface potentials, $rs$ is a good approximation of the sum of the star fractional radii in the case of spherical components.
In Figure \ref{scans} we show the relation between $rs$ and $i$ for the sample of points with a cost function value $<$ 1\% above  the minimum. As expected  the correlation
is very strong.  The projection on the y-axis of the distribution provides an estimate of the uncertainty on $rs$  ($\sim$ 0.001) taking the correlation with $i$ into account. This is 
larger than the formal error on the same quantity, which can be derived from  the LS fit results,  by a factor of $\sim 8$. 

The uncertainties on
the adjusted and  derived parameters reported in Table \ref{sol} correspond to the 1\% threshold (and to the similar BR results),  all significantly larger than the formal errors 
of the LS fits. The  formal errors from the LS fit are typically a factor of ten smaller. The error on the primary temperature, which is fixed in the solution, derives from  the   analysis of the disentangled spectra.

\section{Spectroscopic analysis}
\label{abundances}

\subsection{The effective temperatures}
\label{temps}
Due to Doppler shifts the spectra of binary stars are complex and
vary in the course of the orbital cycle. Usually the spectral lines
of the components are overlapping, especially around eclipses,
what in turn makes the radial velocities difficult to measure and inaccurate.
 The method of {\sl spectral disentangling}, ({\sc spd}, for short), 
overcomes these difficulties \citep{simon1994, had1995}. 
In a self-consistent way it
returns the optimal set of  orbital elements and the individual
spectra of the components. The gain of {\sc spd} is obvious: 
separated spectra 
could be analyzed as  single star spectra, without the interfering
signal of the companion. This allows the spectroscopic determination
of the effective temperatures and of detailed elemental abundances
\citep{hpv2000,pav578,fremat2005,jvc2008,hareter2008,pav453,andrew2009}. 
The effective temperature
determined from hydrogen line profiles, and/or from the  strength of 
temperature sensitive lines, supersede in  accuracy and
precision those derived from e.g.~broad-band or Str\"{o}mgren
photometry. The same is true for the metallicity derived from a
detailed abundance study rather than from photometric colour indices.

For the purpose of spectroscopic diagnostics of the component
spectra {\sc spd} was performed on the time-series of the FEROS
spectra,  since only these spectra cover the hydrogen Balmer lines on which
is based the determination of the effective temperatures.
 We  used {\sc FDBinary}\footnote{\sf
http:sail.hr.fdbinary.html} \citep{ilijic2004} which  performs spectral disentangling
in the Fourier space according to the  prescription of \citet{had1995}. In spite of the very
limited number of spectra {\sc spd} performed quite well and,   thanks to fair phase distribution
of the observations \citep{hensberge2008}, did not produce  strong
undulations in disentangled spectra. 

\begin{figure}
\resizebox{\hsize}{!}{\includegraphics{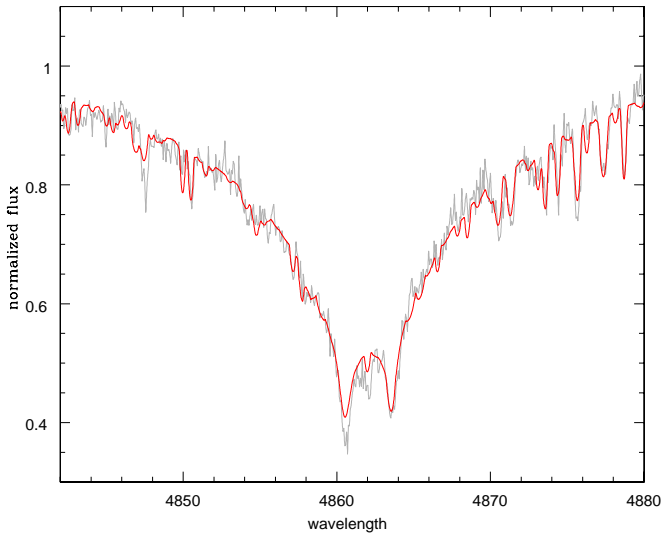}}
\caption{
The best fitting synthetic composite spectrum of CoRoT 102918586
(solid line in dark grey). The spectral window is centered
on the H$\beta$ line. The  FEROS spectrum observed on HJD 2455558.64812
is shown in light grey. The primary lines are blue-warded with respect to 
the secondary ones.}
\label{fbalmer}
\end{figure}

The determination of  the effective temperature from the  profiles
of the hydrogen Balmer lines suffers from degeneracy in \tef\ and
$\log g$, and that is a major limitation in setting-up the appropriate
model atmosphere for a detailed abundance study. However, the complementary
analysis of the light and RVs curves of eclipsing  double-lined
spectroscopic binaries provides as well an independent and accurate 
determination of the  surface gravity of the components. In well-determined
solutions the accuracy in \lgg\ is around 0.01 dex or less, and that is indeed  our case, so
that the degeneracy in \tef\ and \lgg\ could be lifted.  

The components' spectra are diluted in a binary star spectrum proportionally
to their relative light contribution to the total light of the system.
In the observed spectra of a binary system there is no information
on the absolute spectral line strengths and, unless there is at least one
spectrum taken in  eclipse, there is an ambiguity in disentangled
component spectra.  Therefore, a renormalisation of the disentangled spectra
of the individual components to their own continuum shall be performed
using some additional information, as discussed in  detail by
\citet{pav_brno} and \citet{pav_tatre}. 

In the present analysis we used a {\sl constrained optimal fitting}
method \citep{tamajo2011}. 
In the new release of the computer
code for  constrained optimal fitting of disentangled component
spectra the following parameters can be adjusted for both stars:   
 effective temperature, surface gravity,
 light dilution factor, Doppler shift,  projected rotational velocity,
and the vertical shift to adjust for the continuum (in spectral disentangling
the disentangled component spectra are shifted and an additive factor shall
be applied to return them to the continuum level \citep{ilijic2004,pav578, andrew2009}.
 
The disentangled spectra were then analyzed to derive the atmospheric parameters of 
each component. 
To this purpose a comprehensive grid of theoretical spectra was calculated for$T_{\rm eff}  = 5000$ to 15000 K,
and \lgg = 2.5 to 5.0 in steps of 250 K in  temperature and 0.5 in  \lgg .
The {\sc atlas9} model atmospheres of \citet{atlas9} and the {\sc uclsyn} spectral
synthesis package 
\citep{uclsyn} were used for the construction of the grid  ( which is for fixed solar metallicity
but  was used only in the initial determination of \teff\  from the Balmer lines, see below).
 A genetic algorithm was then used \citep{charb} for the optimization with respect to  model parameters.
Further detailed description of the code is given in Pavlovski et~al.\ (2012, in preparation).    

Only the hydrogen Balmer lines  H$\gamma$ and H$\beta$ were selected for the optimal
fitting, excluding from the fit
 the blends due to the various metal lines. The surface
gravities and the light factors of both components were fixed according to the light curve
solution (Table~\ref{sol}). 
Since the Balmer lines are almost insensitive to 
rotational broadening, with the exception of the very core of the lines, 
the projected rotational velocities were also kept fixed.
Their values were derived from the clean
metal lines, in an iterative cycle with the determination of the effective
temperatures from Balmer lines yielding  $v_{\mathrm{rot}} \sin i = 20.5 \pm 0.5$ \ks
and $v_{\mathrm{rot}} \sin i = 14.5 \pm 0.5$ \ks for the primary and secondary component,
respectively.
With the surface gravities, the light dilution factors and  projected 
rotational velocities for both components fixed, the only free parameters in the
the optimal fitting were the effective temperatures, the Doppler shifts of
disentangled spectra to reference frame of the synthetic spectra, and
the vertical adjustment of the continuum level for both components
separately. 

The mean values
of the effective temperatures derived in these calculations,
separately for two Balmer lines read: \teff\ = 7\,400$\pm$90 K for
the primary, and \teff\ = 7\,100$\pm$110 K for the secondary component. The
latter is in excellent agreement with the value derived from the light curve solution in  Table \ref{sol}. 
The quality of the fit for the composite H$\beta$ line in one of the
observed spectra is shown in
Fig.~\ref{fbalmer}.

\begin{figure}
\resizebox{\hsize}{!}{\includegraphics{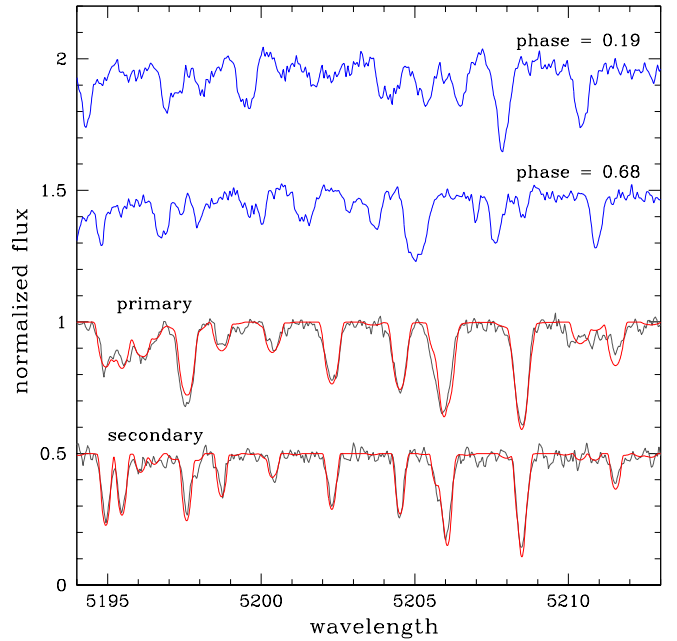}}
\caption{
Short spectral stretch of two observed composite spectra at different
orbital phases from the FEROS dataset. The two individual disentangled
spectra are shown at the bottom along the best fitting synthetic
spectra calculated for the effective temperatures and surface
gravities derived in Sect. \ref{lc-rv-an} \& \ref{abundances} (Tab. \ref{sol} \& \ref{atmos}).
}
\label{spectra}
\end{figure}

\begin{table}[b!]
 \centering
\caption{Atmospheric parameters  of  CoRoT~102918586 components$^{\mathrm{a}}$.  }
\label{atmos}
\begin{tabular}{lcc}
\hline
Parameter                             &  Star A       & Star B        \\
\hline
$T_{\rm eff}$ (K)    &   $7380\pm90$  &  $7100 \pm110$   \\
$v_{\mathrm{rot}} \sin i$ (\ks)    &   $20.5\pm0.5$     &    $14.5\pm0.9$    \\
$\xi_{\rm t}$  (\ks) &   $1.8\pm0.1$    &    $1.0\pm0.1$   \\
\hline
\end{tabular}
\begin{list}{}{}
\item[$^{\mathrm{a}}$] {\teff's are derived from the Balmer lines, \vsin\ and  $\xi_{\rm t}$
from metallic lines.}
\end{list}

\end{table}

\subsection{Abundances}

The important outcome of spectral disentangling is the gain in S/N ratio.
That is evident  from the short section of the  component disentangled spectra 
 which are displayed in the lower part of Fig. \ref{spectra}, together with two composite 
 observed spectra at different orbital phases. 
 
We estimated an enhancement in S/N, thanks to spectral disentangling and binning of disentangled spectra, 
by  the factor of about  2.2 and 1.6, for
the primary and the secondary respectively. The difference is  due to the unequal contribution of the
components to the  total light.
The increase in S/N  makes, in turn, possible the abundance study. 

First, we estimated the microturbulent velocity $\xi_t$. In the spectra
of F-type stars \ion{Fe}{i} lines serve to the purpose, since they are the
most numerous. 
The mictroturbulent velocities for the primary and secondary component,
minimizing the scatter in the abundances from different iron lines,
are  $\xi_{\rm t,p} = 1.8\pm0.1$ km\,s$^{-1}$ and $\xi_{\rm t,s} = 
1.0\pm0.1$ km\,s$^{-1}$.

 Iron appears in two ionization stages, \ion{Fe}{i} and \ion{Fe}{ii},
which were used as additional diagnostics of the effective temperatures
\citep{barry2005}. The null-dependence of the iron abundance on the excitation
potential slightly corrected the \tef's, in a larger amount  for the secondary component. 
However, the \ion{Fe}{i} lines outnumber the \ion{Fe}{ii} lines, and the solution is
not well constrained. Differences of about 0.20 dex are found in the iron abundances
from  \ion{Fe}{i} and \ion{Fe}{ii}, for the primary and secondary component,
which are larger than the estimated errors in abundance determination. In both
cases the iron abundances are larger from the \ion{Fe}{ii} lines. Similar differences
were found also for other ions which appear in two ionization stages.
 However,  the \ion{Fe}{i}
lines outnumber the \ion{Fe}{ii} ones by factor of 5 and 6 for the primary
and secondary component, respectively, and we eventually adopted abundances
derived from \ion{Fe}{i} lines as  final. Still, this enables a slight 
correction of \tef's derived in the previous subsection, and the final, 
adopted \tef's are listed in Table~\ref{atmos}. The final adopted elemental abundances, 
as derived in LTE approximation, are 
listed in Table~\ref{abund}.
The abundances relative to hydrogen are given in the second and  fourth columns ($\log  N({\rm
H})=12.00$),  those relative to the solar one   in the third and fifth columns, with 
the solar abundances as given by \cite{gn93}.
 We conclude
that within 1$\sigma$ errors the abundances for both components are close
to the solar one. The Iron abundances [Fe/H]$_{\rm p}$ = 0.11$\pm$0.05 and 
[Fe/H]$_{\rm s}$ = -0.10$\pm$0.04 give an approximate metallicity
for the components, $Z_{\rm p} = 0.020 \pm 0.001$ and $Z_{\rm s} = 0.016 \pm 0.001$,
being $Z_{\odot} = 0.02$ from  \cite{gn93}. Because of the uncertainties in the light ratio
between the components, which in turn could affect our \tef's determination,
we conclude that the metallicity of both components can be assumed to be
solar.

 \begin{table}[t!]
 \centering
\caption{\label{abund}Photospheric elemental  abundances of
CoRoT~102918586.}
\begin{tabular}{ccccc}
\hline
Ion            &  $\epsilon(X)_{\rm p}$ & [X/H]$_{\rm p}$ &
$\epsilon(X)_{\rm s}$ & [X/H]$_{\rm s}$   \\
\hline
\ion{C}{}      &    -           & -               &  8.51$\pm$0.09  & -0.04$\pm$0.10    \\
\ion{Si}{}     &  7.43$\pm$0.05 & -0.12$\pm$0.07  &  7.48$\pm$0.06  & -0.07$\pm$0.08    \\
\ion{Ca}{}     &  6.22$\pm$0.15 & -0.14$\pm$0.15  &  6.06$\pm$0.14  & -0.30$\pm$0.14    \\
\ion{Ti}{}     &  5.11$\pm$0.17 & 0.09$\pm$0.18   &  4.94$\pm$0.09  & -0.08$\pm$0.11    \\
\ion{Cr}{}     &  5.83$\pm$0.08 & 0.16$\pm$0.09   &  5.54$\pm$0.12  & -0.13$\pm$0.12    \\
\ion{Mn}{}     &  5.55$\pm$0.11 & 0.16$\pm$0.11   &  5.54$\pm$0.09  &  0.15$\pm$0.09    \\
\ion{Fe}{}     &  7.61$\pm$0.03 & 0.11$\pm$0.05   &  7.40$\pm$0.03  & -0.10$\pm$0.05    \\
\ion{Ni}{}     &  6.37$\pm$0.08 & 0.12$\pm$0.09   &  6.23$\pm$0.07  & -0.02$\pm$0.08    \\
\hline
\end{tabular}
\end{table}
    \begin{figure}[hb!]
   \centering
   \includegraphics[width=8.7cm]{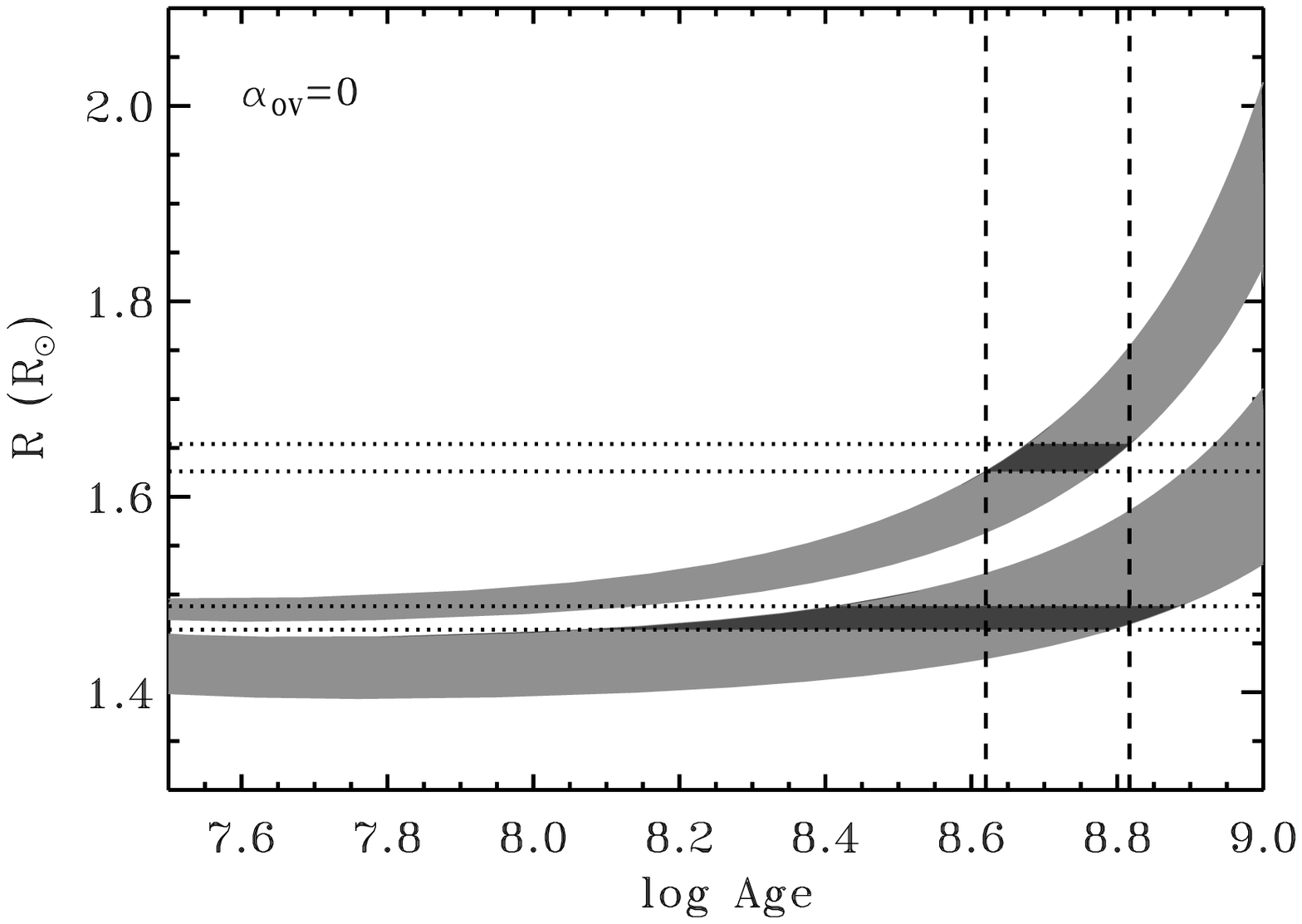}
   \includegraphics[width=8.7cm]{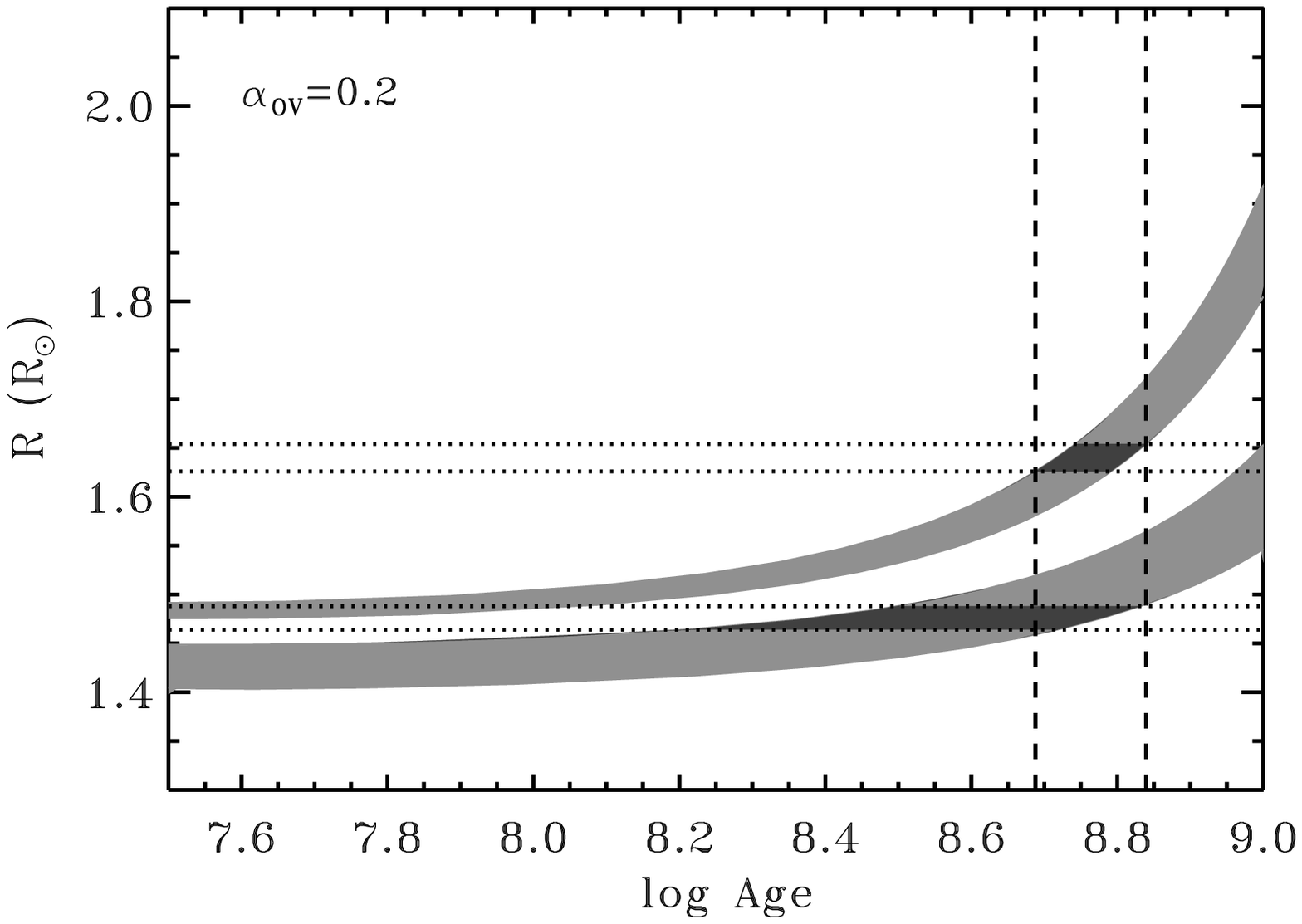}
   \caption{Evolution of the component radii. Upper panel: based on models without overshooting, $\alpha_{{\mathrm OV}}=0$, lower panel: $\alpha_{{\mathrm OV}}=0.2$ . The light-gray bands describe the time evolution of the radii for star in the mass range  $M_{1,2} \pm \Delta M_{1,2}$, from Table~\ref{sol}. The darker areas fill the intersection with the $R_{1,2} \pm \Delta R_{1,2}$ lines from the same Table. The vertical lines bound the time interval respecting the constraint of coevality. }
         \label{radev}
   \end{figure}
     \begin{figure}[htb!]
   \centering
   \includegraphics[width=9.cm]{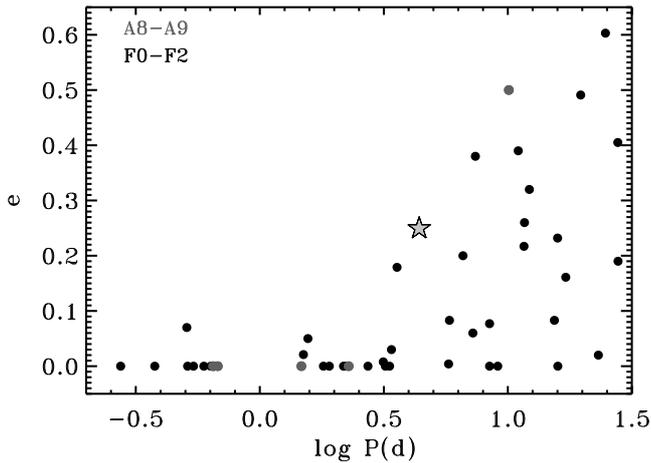}
   \caption{The location of \eb (starred symbol) in the period - eccentricity diagram for stars of spectral type A8-F2   \citep[derived from the database  of the  9$^{{\rm th}}$ Catalogue of Spectroscopic Binary Orbits, ][]{sb9}}
         \label{pecc}
   \end{figure}
\section{Physical properties of \eb} 
\label{abspar}

A comparison with theoretical models of the given mass and chemical  composition provides further information on the physical properties of the system.
For instance, the inspection of the evolution of the stellar radii with time, combined to the constraint of coevality, allows an accurate determination of the system age.
We prefer to use the  evolution of the radius, instead, for instance, of the position in the Hertzsprung--Russell (H-R)  diagram, because the radii are directly  and precisely determined from the radial velocity and light curve  solutions, while the computation of temperature and luminosity  implies the use of color transformations and bolometric corrections. 

The evolutionary tracks shown in Fig.~\ref{radev}  were obtained from stellar evolution modeling with the code CLES
 \citep[Code Li\'{e}geois  d'\'{E}volution Stellaire,][]{scuetal08}. 
   The computations were made with the   equation of state from OPAL \citep[OPAL05,][]{ronay02}. The opacity tables are, as well,  from OPAL \citep{iglrog96} for the solar mixture of \citet[][GN93]{gn93}, extended to low temperatures with the  \citet{fergal05} opacity values. The nuclear reaction rates are those of the NACRE compilation \citep{angulo99} except for 
 $\mathrm{ ^{14}N(p,\gamma)^{15}O}$, updated by \citet{formi04}.   
  
Models with and without overshooting  were computed with  the mixing-length theory (MLT) of convection \citep{bv58}.  The value of  $\alpha_{\rm MLT}$ was kept fixed, adopting  the solar value of 1.8 (at any rate, in the temperature range of interest here, its value does not affect the stellar radius). 
For the models with overshooting, the overshoot length was expressed in terms of the local pressure scale height  $H_p$, as  $\alpha_{{\mathrm OV}} H_p$,  and the  $\alpha_{{\mathrm OV}}$ value was fixed to $\alpha_{{\mathrm OV}}=0.2$. 
The  chemical composition was assumed to be solar, according to the results of the previous Section. The computations were done  without including microscopic diffusion.

  Figure \ref{radev} shows, for both components,  the region of  the $R\,(\log t)$ diagram  bounded by the evolutionary tracks of mass $M \pm \Delta M$ from Table~\ref{sol}. 
The intersection with the lines  $R \pm \Delta R$ of the same Table provides the possible age range for each star,  the further constraint of coevality narrows the range to the final value  of $540 \pm 120$ Myr, in the case of no-overshooting,  and to $580 \pm 100$ Myr for  $\alpha_{{\mathrm OV}}=0.2$. In both cases a young system with two Main Sequence (MS) stars which have burnt less than 20\% of their initial hydrogen content.

Because of the young age the binary had the time reach only  spin-orbit synchronization while orbit circularization is still on the way. The eccentricity is quite high for the orbital period, as  can be  seen  (Fig.~\ref{pecc}), from the location of \eb  in the period-eccentricity diagram of stars of similar spectral type.  For periods smaller than $\sim$10 days the circularization process is expected  to be efficient enough to change the system eccentricity in a few Gyr. So  in Fig. \ref{pecc}  the younger stars  born with high eccentricity  trace the upper envelope of the distribution, an will evolve with time towards lower and lower values.

\section{Pulsational properties} 

\label{puls}
{
The final frequency spectrum, after subtraction of the binary model, is shown in Fig.~\ref{frsp}.  The plot is restricted to the frequency interval containing meaningful features as, for higher frequencies, we only detected signal related to the satellite orbit.  

 As mentioned before, to evaluate the frequency significance  we used the criterion suggested by  \cite{breger93}, i.e. a S/N value of the amplitude  $\ge$ 4. The   twenty-eight frequencies satisfying this constraint are listed in Table~\ref{hff}.  The errors in the table are derived according
to \cite{kallinger08}  and are sensibly larger that the formal errors of the LS fit, the remark column lists the closest frequency combinations, taking the uncertainties into account. 

 The first three dominant frequencies of the amplitude spectrum are separated by $\sim0.05$ d$^{-1 }$. The fourth frequency is the beating   of the third highest amplitude frequency (F3 of Table~\ref{hff}) and F$_{\mathrm{orb}}=0.22772$ \cd. Other combinations of these  four highest amplitude frequencies are also clearly visible. Figure~\ref{frsp} shows as well the window DFT,  whose main features are the peaks at 13.972 \cd and its day aliases, due to the satellite orbital frequency, and a peak at  2.005 \cd due to the South Atlantic Anomaly crossing.  The gray line  is the significance threshold which is four times the local mean of the residual spectrum.

  The F$_{\mathrm{orb}}$ overtones present in the initial analysis of Table~\ref{hf0} have disappeared or very weak, which is an indirect confirmation of the successful removal of the orbit-only binary light curve.  The low frequency components (F $< 0.1$ \cd), are presumably leftovers of the detrending procedure with a low order polynomial.
\begin{figure}[htb]
   \centering
   \includegraphics[width=9.cm]{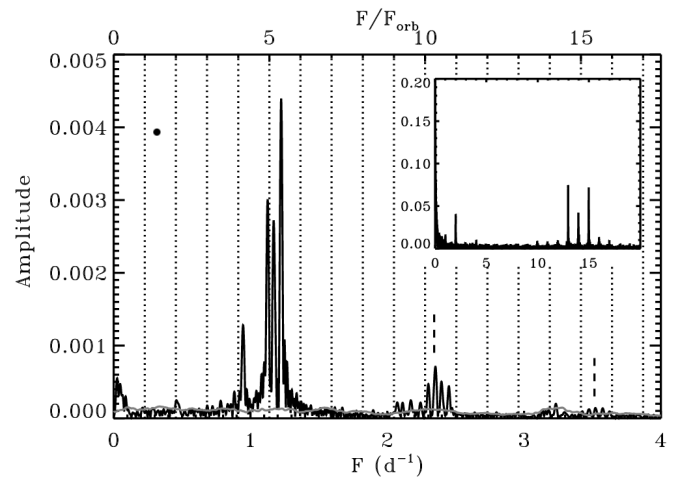}
   \caption{The final amplitude spectrum of \eb and, in the inset, the window DFT.  The vertical dotted lines correspond to the orbital period overtones, the broken lines to 2F3 and 3F3. The gray line is the threshold of significance (S/N$\ge 4$) computed as explained in the text. }
         \label{frsp}
   \end{figure}
\begin{table*}[ht!]
 \centering
\caption{The frequencies derived from the analysis of the final set of residuals, after 
subtraction of the EB model, with an amplitude S/N$\ge$ 4.} 
\label{hff}
\begin{tabular}{llllll}
\hline
       &  F (\cd)             & Ampl $\cdot 10 ^3$ &  Phase (2$\pi$)      & S/N& remark $^{\mathrm{a}}$\\   
\hline  
F1  & 1.2247 $\pm$  0.0017 &  4.25 $\pm$  0.41 &  0.59  $\pm$ 0.04 & 17.8  &   g \\
F2	& 1.1255 $\pm$  0.0012 &  3.10 $\pm$  0.22 &  0.16  $\pm$ 0.03 & 24.5  &   g \\
F3	& 1.1716 $\pm$  0.0010 &  2.73 $\pm$  0.16 &  0.03  $\pm$ 0.03 & 29.3  &   g \\
F4	& 0.9462 $\pm$  0.0017 &  1.42 $\pm$  0.14 &  0.19  $\pm$ 0.05 & 17.2  &   F3-F$_{\mathrm{orb}}$\\
F5	& 2.3520 $\pm$  0.0022 &  0.72 $\pm$  0.09 &  0.01  $\pm$ 0.06 & 13.6  &   F1+F2  \\
F6	& 0.0252 $\pm$  0.0032 &  0.67 $\pm$  0.13 &  0.41  $\pm$ 0.09 & 9.2   &   lf	    \\
F7	& 2.3964 $\pm$  0.0026 &  0.51 $\pm$  0.07 &  0.48  $\pm$ 0.07 & 11.6  &   F1+F3  \\
F8	& 2.4486 $\pm$  0.0027 &  0.42 $\pm$  0.07 &  0.24  $\pm$ 0.07 & 11.0  &   2F1	\\
F9	& 2.3015 $\pm$  0.0026 &  0.42 $\pm$  0.06 &  0.53  $\pm$ 0.07 & 11.2  &   F2+F3  \\
F10	& 0.0566 $\pm$  0.0055 &  0.45 $\pm$  0.14 &  0.17  $\pm$ 0.15 & 5.4   &   F1-F3 	\\
F11	& 1.2117 $\pm$  0.0045 &  0.36 $\pm$  0.09 &  0.67  $\pm$ 0.12 & 6.6   &   F1\\
F12	& 0.0409 $\pm$  0.0050 &  0.28 $\pm$  0.08 &  0.97  $\pm$ 0.13 & 5.9   &   lf  \\
F13	& 2.1709 $\pm$  0.0035 &  0.30 $\pm$  0.06 &  0.49  $\pm$ 0.09 & 8.5   & F1+F3-F$_{\mathrm{orb}}$ \\
F14	& 0.4666 $\pm$  0.0063 &  0.29 $\pm$  0.10 &  0.96  $\pm$ 0.17 & 4.8   & 2 F$_{\mathrm{orb}}$ \\
F15	& 1.1133 $\pm$  0.0047 &  0.26 $\pm$  0.07 &  0.65  $\pm$ 0.13 & 6.4   & 2F3-F1\\
F16	& 2.0726 $\pm$  0.0043 &  0.25 $\pm$  0.06 &  0.77  $\pm$ 0.11 & 7.0   & F2+F3-F$_{\mathrm{orb}}$ \\
F17	& 2.1161 $\pm$  0.0041 &  0.23 $\pm$  0.05 &  0.52  $\pm$ 0.11 & 7.3   & F1+F2-F$_{\mathrm{orb}}$\\
F18	& 1.1925 $\pm$  0.0055 &  0.21 $\pm$  0.07 &  0.36  $\pm$ 0.15 & 5.4   & F3-F6\\ 
F20	& 1.2744 $\pm$  0.0061 &  0.16 $\pm$  0.06 &  0.69  $\pm$ 0.16 & 4.9   & 2F1-F3\\
F21	& 2.2493 $\pm$  0.0045 &  0.15 $\pm$  0.04 &  0.67  $\pm$ 0.12 & 6.6   & 2F2 \\
F22	& 3.5254 $\pm$  0.0041 &  0.15 $\pm$  0.03 &  0.27  $\pm$ 0.11 & 7.3   & F1+F2+F3 \\
F23	& 0.0113 $\pm$  0.0046 &  0.24 $\pm$  0.06 &  0.20  $\pm$ 0.12 & 6.5   & lf\\
F24	& 3.5750 $\pm$  0.0043 &  0.13 $\pm$  0.03 &  0.03  $\pm$ 0.11 & 7.0   & F1+2F3 \\
F25	& 3.4758 $\pm$  0.0044 &  0.12 $\pm$  0.03 &  0.54  $\pm$ 0.12 & 6.7   & F2+2F3 \\
F26	& 1.1394 $\pm$  0.0065 &  0.15 $\pm$  0.06 &  0.72  $\pm$ 0.18 & 4.6   & F1-2F2-2F3\\
F27	& 3.6220 $\pm$  0.0047 &  0.10 $\pm$  0.03 &  0.64  $\pm$ 0.13 & 6.3   & 2F1+F3 \\
F28 & 3.2991 $\pm$  0.0060 &  0.09 $\pm$  0.03 &  0.60  $\pm$ 0.16 & 5.0   & F1+F2+F3-F$_{\mathrm{orb}}$\\
\hline 
\end{tabular}
\begin{list}{}{}
\item[$^{\mathrm{a}}$] {Closest frequency combination. The independent frequencies are denoted with ``g", the low frequency component with ``lf".}
\end{list}

\end{table*}
   
An oscillatory pattern of the kind  superimposed on our EB light curve  is not necessarily due to pulsations. A study  of the large sample of variables detected by the {\it Kepler} satellite and classified as $\gamma$~Dor   \citep{bal-gd} has shown that the variability of ``symmetric"  type, i.e symmetric with respect to its mean value,  may  be due  to  stellar spots experiencing differential rotation, and that  the {\it Kepler} sample of $\gamma$~Dor is contaminated by non-pulsating star. In this case the dominant frequency is that of stellar rotation and the beating is due to slightly different rotation rates of  migrating spots.

In our case, however, even if the oscillatory pattern does show symmetry with respect to its mean value, we can exclude this hypothesis, because  spectroscopy allowed  an accurate measurement of $v_{\mathrm{rot}} \sin i$ (and  we have from the light curve analysis  the value of the inclination). Both stars have rotation frequencies  close to the orbital one (1.1 and 0.9  F$_{\mathrm{orb}}$ for the primary and secondary, respectively), and  F$_{\mathrm{orb}} \simeq$ F$_{\mathrm{rot}}$ is much smaller than the dominant frequency of the  amplitude spectrum.
 If the rotation axes were not aligned with the orbital one, the stellar rotation frequencies could be higher, but the disalignement  should be conspicuous: the factor  $\simeq 5$ between the dominant  frequencies and F$_{\mathrm{orb}}$   requires an inclination of the rotation axis as low as 10$^\circ$.  Besides, the similarity between the rotation frequencies  -- as derived in the hypothesis of alignment -- and the orbital frequency would remain   an  unexplained coincidence.

We adopt, therefore, the simplest scenario in which the periodic variations are not due to rotation but to stellar pulsations. 
     \begin{figure}[htb]
   \centering
   \includegraphics[width=9.cm]{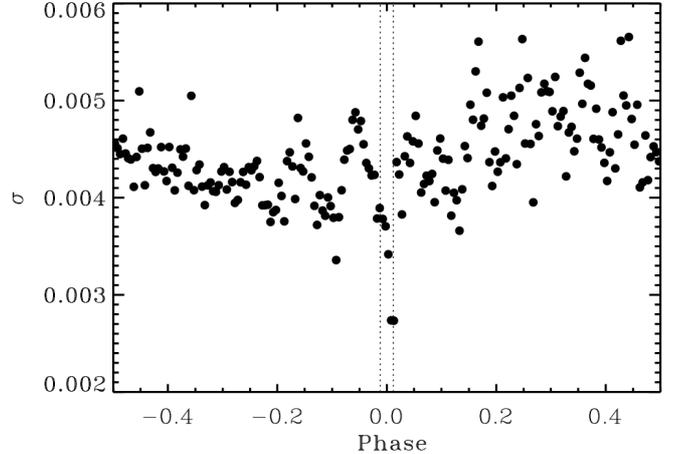}
   \caption{The standard deviation, with respect to the local mean  of the binned residual light curve (see text).  The vertical lines bound the in-eclipse phases. The  dip around phase zero suggests that the  pulsations  belong to the eclipsed component, i.e. to the primary star.}
         \label{sigmavar}
   \end{figure}
 The dominant frequencies seen in the spectrum (F1 -- F3)  fall indeed in the range of excited high order g-modes typical of  $\gamma$ Dor pulsators. 

We have, therefore, computed with MAD \citep{MAD03} the instability domain for  stellar models  with the physical characteristics of the  components.
In particular, in the case of models without overshooting  the range is  0.50 -- 4.16  ($\ell=1$)  and 0.52 -- 5.52  \cd ($\ell=2$) for the primary, and 0.55 -- 4.7 ($\ell=1$) and  0.55 -- 5.42 \cd ($\ell=2$)
for the secondary component. 
The values  change only slightly  for the models with overshooting, becoming  0.51 -- 3.83  ($\ell=1$) and 0.55 -- 5.02 \cd ($\ell=2$)   for the primary,  and 0.55 -- 4.16  ($\ell=1$) and 0.53 -- 5.33 \cd ($\ell=2$) for the companion.  Being the range very similar for both stars  it is impossible to identify  the  pulsator in this way. 

We have found, however, a possible indirect evidence that the pulsating star is the primary, or at least that the higher amplitude frequencies belong to the primary.  This 
might be, in fact, the reason why the  dispersion of the phased light curve decreases during  eclipse.  To show this effect we re-binned the residual light curve (after subtraction of the best binary model) in 200 equal  phase-bins and  computed the local  standard deviation, which is shown in Fig. \ref{sigmavar} as function of the orbital phase. There is an evident dip  around phase zero.
However, taking into account that  we have only thirteen (narrow)  minima in the whole light curve, implying a  poor sampling of  pulsation in eclipse, and
that only a small fraction of the primary surface is out of view during a grazing eclipse, we think that this conclusion shall be taken with caution.

 Most  frequencies shown in Table~\ref{hff} can be re-conducted to a combination of the first three ones and the orbital frequency.  
 The main features in the spectrum are three groups centered around  F3, F5 and F22 (1.1716, 2.3520 and 3.5254 \cd, respectively).  Each of those frequencies is the center of a structure with side peaks spaced by about 0.05 \cd.  
 This fact could suggest an interpretation in terms of rotational splitting of the central frequency. In particular  both F5 and F22 have two peaks each side  (one in the case of F22 just  below the significance threshold,  Fig \ref{frsp}). This occurence could be interpreted --  at first sight --  as  quintuplets due the effect of rotation of an $\ell=2$ mode, which are  considered the ones preferentially excited by tidal forces \citep[e.g.,][]{koso83}.
 
 There are, however, several arguments against this hypothesis.
 The first is, again,  the known values of the component rotation.  As the star is rotating slowly enough  (its rotation frequency is  1.5 \% of the critical value
 as defined by \citet{tow05}  and smaller than the dominant pulsation frequencies),   the splitting can be expressed according to the   first order approximation  by \cite{ledoux51}:  
\begin{equation}
\label{led51}
\Delta F=\left ( 1- \frac{1}{\ell(\ell+1)} \right )F_{\mathrm{rot}} 
\end{equation}
\noindent where $\ell$ is the degree of the mode and $ F_{\mathrm{rot}} =0.253$ \cd the rotation frequency. In our case, with $\ell=2$,  $\Delta F \simeq 0.21$ \cd, i.e. about four times the  measured value.  In the case of a dipolar mode,  $\ell=1$, the resulting splitting of the corresponding  triplet would be smaller but  still   $\sim$2.5 times larger than the derived
value.   

Even if we interpret the measured value as reflecting the average value of the stellar rotation (which is probably not rigid), we would end up with a star whose surface (observed) rotation rate is much faster than that of its interior.
That might be the case of late-type (solar and later) components  in close binaries, which might be spun up starting from a slow rotation regime as a consequence of  angular momentum loss by magnetic braking and spin-orbit synchronization by tidal action. Depending on the core-surface coupling, the external layers could, in this case, rotate faster than the interior.  The components of \eb, however, are early F stars, so the previous scenario  does not apply.

Besides, it has to be noticed that  the actual value of the spacings F3-F2 and F1-F3 are not exactly 0.05 but rather   $0.0461 \pm 0.0016$ and $0.0531 \pm 0.0020$, respectively and, again on the basis of the slow rotation, one would expect symmetrical peaks.  

Last, but not least, there exists an alternative and more convincing interpretation of the central frequencies as combinations: F5 is closer to F1+F2 than to 2F3 and F22 to F1+F2+F3.  The remarks in  Table~\ref{hff} correspond to the best matching combinations. So we can safely conclude that the patterns in Fig. \ref{frsp} are not related to stellar rotation.

The comparison of the frequency spectrum with those of other $\gamma$~Dor stars, observed by \corot and  {\it Kepler}, suggests that the pattern we see is quite common  in (presumably) single stars as well \citep[see, e.g.,  ][]{bal-gd, hareteral10}. For instance, the light curve\footnote{The light curve  and its plot are available  from the public section of the CoRoT-N2 archive at  the CoRoT Data Center (IAS): http://idoc-corot.ias.u-psud.fr/ } of  CoRoT~102732872 (an F3V star, pulsating with a
main frequency of 1.09 \cd) closely resembles that of \eb after removal of the EB contribution.   The former star is one of the few $\gamma$~Dor which,  according to \cite{hareteral10},  shows a pattern of regularly spaced peaks in period (as expected in the case of high-order g-modes).  We computed, therefore,   the mean period spacing of the three independent frequencies, $\Delta P=0\fd0360 \pm0\fd0011$ (or 3110$^\mathrm{s} \pm 90^\mathrm{s} $) and compared it with the  theoretical asymptotic period spacing of the $\ell=1$ high-order g-modes, as expected for the stellar parameters from Table~\ref{sol} and Section \ref{abspar}.

 The computed range of period spacing for the primary star, relative to the stellar models in the age interval derived in Section~\ref{abspar},  is 3070-3200$^{\mathrm{s}}$ for the models without overshooting, and 3300-3390$^{\mathrm{s}}$ for those with  $\alpha_{\mathrm{OV}}=0.2$. For the secondary star the values are, respectively
2800- 3030$^{\mathrm{s}}$ and 3160 - 3300$^{\mathrm{s}}$. A consistent interpretation is, therefore, that the observed pattern is due to non-radial  dipolar  pulsations  of the primary component.

\section{Discussion and conclusions} 

\eb is the first $\gamma$ Dor in an eclipsing binary for which it was possible to collect observations of such a quality to allow a very detailed study,which provided both an accurate determination of the physical parameters  and of the pulsational properties.  Masses and radii have uncertainties $\le 1$\%,  the effective temperatures of 1.5 \% and the derived age,  turned out to be  600 Myr, within a range of $\sim$ 100 Myr.

Our results   depict the system as a young eccentric binary  formed by two similar MS star (the mass ratio is 0.9), the single eclipse is just due to the orientation of the binary orbit in space and the primary component is at the origin of the observed $\gamma$~Dor pulsations. We cannot completely exclude that the similar secondary component pulsates as well, but we suggest the dominant pulsation frequencies  belong to the primary, on the basis of the  measured  decrease of the pulsation amplitude during the eclipse.  

 Spectra disentangling of the FEROS spectra allowed to derive the effective temperatures and the component chemical abundances, which turned out to be close to solar. The abundance pattern for the $\gamma$ Dor pulsating stars was studied by \cite{Brunttal08}, and  more recently by \cite{Tkach12}, particularly in regard to the links with the chemically peculiar $\lambda$~Boo and Am-type stars suggested by \cite{gray99} and \cite{Sadak06}. Their comprehensive detailed abundance analysis  did not reveal any sign of  chemical
peculiarities  in their sample of $\gamma$ Dor stars, yielding the conclusion that the metallicity is quite close to the solar value. Previously, on the  basis of the
Str\"{o}mgren photometric $m_1$ index, \cite{hand99} as well noted that all the $\gamma$ Dor candidates in his sample had metallicities close to
the solar value. Our finding for the $\gamma$ Dor pulsating component in the binary system CoRoT 102918586 corroborates this general conclusion.

The results of our asteroseismic study are limited by the lack of calibrated multicolor photometry  in CoRoT and/or  of spectroscopy of sufficiently high S/N to allow to study line profile variations due to pulsation (unfeasible given the system magnitude). As a consequence, we could not completely identify the pulsation modes. On the other hand, from the comparison of the period splitting with the theoretically expected values, we have at least indications that the primary component is pulsating in a $\ell =1$ mode. Those are not the modes preferentially excited by  tidal forces, which have been detected (or suggested) in  in other cases. \cite{handl02} propose that a number of frequencies, exact multiple of the orbital frequency, found in the $\gamma$ Dor component of  HD~209295 (a single-lined spectroscopic binary) could be triggered by tidal interaction.  Another case is that of HD~174884 \citep{cm09}, an eccentric eclipsing binary formed by two B-type stars.

 In \eb, however, we find little or no influence of binarity on the pulsations: the pulsation pattern is very similar to that of single $\gamma$ Dor's and the only sign of the orbital motion is found in the beating of the orbital frequency with the dominant ones.    This might be due to a weak tidal interaction: our system is less eccentric than the above-mentioned ones, has a longer period,  and the components have smaller fractional radii (by a factor of $\sim 1/2$ with respect to HD 174884 and even smaller in the case of HD~209295, whose components radii -- however -- can only be estimated).  One expects, therefore, a weaker tidal torque, as this is very strongly dependent on the star fractional radius \citep[e.g., ][]{zahn05}.

The final model of the system, characterized by an internal consistency among the results from different data set and treatments,  proves the success of the iterative method used to separate pulsations from the eclipsing binary light curve. That was not obvious, and we do not think that this result is generally valid, because eclipses certainly  modify the observed oscillatory pattern, while we subtracted the same oscillations at all phases.  A reason of the success in this case is that we dealt with a grazing eclipse,  which decreases the determinacy of the light curve solution but  makes this particular problem less severe.

 A straightforward, first-order, correction could be to  weight the pulsation amplitude with the contribution to the total light of the pulsating component during eclipses, but  the definite solution is the inclusion of non-radial pulsation in the binary model, taking into account  the deformation of the surfaces by non-radial pulsations and the consequent brightness variation. That is however a challenging task, as we do not know a priori in which modes the star oscillates, and adding the parameters governing the oscillations to the minimization procedure  greatly increases its complexity (and its unicity problems).
 \begin{acknowledgements}
This work is partly based on public CoRoT data and we  are very grateful to the CoRoT scientific team for  making available the mission results in the \corot archives.

 We thank   Artie Hatzes, P.I. of the echelle-spectroscopy program at McDonald Observatory, for including our binary in their list of CoRoT exoplanet  targets,  Jonas
 Debosscher for providing the results of the classification of CoRoT variables, and  Hans Deeg for checking the level of contamination of our
 target.
 We would also like  to express our gratitude to  Andrej Pr\v sa  for making publicly available and maintaining  PHOEBE (and for his constant, patient, support) and to
 Marc Antoine Dupret for making available his code MAD.
  Finally, we  thank   Markus Hareter, Conny  Aerts and  Maryline Briquet for suggestions and fruitful discussions, and the unknown referee for constructive criticisms.

 We acknowledge the generous financial support  by:
 the Istituto Nazionale di Astrofisica (INAF)  under  PRIN-2010   {\it Asteroseismology: looking inside the stars with space- and ground-based
observations}  (CM, MR) and  the Agenzia Spaziale Italiana (ASI) in the frame of the ESS program (CM);
the Belgian PRODEX Office under contract C90199 {\it CoRoT Data Exploitation}  (JM) ;  the German
 DLR under grant 50OW0204 (DG);  the Croatian MZOS under research grant 119-0000000-3135 (KP).

This research has made use of the Exo-Dat database, operated at LAM-OAMP, Marseille, France, on behalf of the CoRoT/Exoplanet program and  of the SIMBAD database,
operated at CDS, Strasbourg, France.
\end{acknowledgements}
\bibliographystyle{aa} 
\bibliography{pebs3} 

 \end{document}